\documentclass{PoS}
\usepackage{cite}

\newcommand{\bq}{\begin{eqnarray}}
\newcommand{\eq}{\end{eqnarray}}
\newcommand{\eps}{\varepsilon}

\title{Differential equations for Feynman integrals beyond multiple polylogarithms}

\ShortTitle{Differential equations for Feynman integrals beyond multiple polylogarithms}

\author{Luise Adams\\
        Johannes Gutenberg-Universit\"at Mainz\\
        E-mail: \email{ladams01@uni-mainz.de}}

\author{Christian Bogner\\
        Humboldt-Universit\"at zu Berlin\\
        E-mail: \email{bogner@math.hu-berlin.de}}

\author{Ekta Chaubey\\
        Johannes Gutenberg-Universit\"at Mainz\\
        E-mail: \email{ladams01@uni-mainz.de}}

\author{Armin Schweitzer\\
        ETH Z\"urich\\
        E-mail: \email{armin.schweitzer@phys.ethz.ch}}

\author{\speaker{Stefan Weinzierl}\\
        Johannes Gutenberg-Universit\"at Mainz\\
        E-mail: \email{weinzierl@uni-mainz.de}}

\abstract{
Differential equations are a powerful tool to tackle Feynman integrals.
In this talk we discuss recent progress, where the method of differential equations has been
applied to Feynman integrals which are not expressible in terms of multiple polylogarithms.
}

\FullConference{13th International Symposium on Radiative Corrections (Applications of Quantum Field Theory to Phenomenology)\\
		 25-29 September, 2017 \\
		 St. Gilgen, Austria}

\begin{document}

% -----------------------------------------------------------------------------

\section{Review of differential equations and multiple polylogarithms}

The method of differential equations \cite{Kotikov:1990kg,Kotikov:1991pm,Remiddi:1997ny,Gehrmann:1999as,Argeri:2007up,MullerStach:2012mp,Henn:2013pwa,Henn:2014qga,Ablinger:2015tua,Bosma:2017hrk}
is a powerful tool to tackle Feynman integrals.
Let $t$ be an external invariant (e.g. $t=(p_i+p_j)^2$) or an internal mass and let $I_i \in \{I_1,...,I_N\}$ be a master integral.
Carrying out the derivative $\partial I_i/\partial t$ 
under the integral sign and using integration-by-parts identities allows us to express the 
derivative as a linear combination of the master integrals:
\bq
 \frac{\partial}{\partial t} I_i
 & = &
 \sum\limits_{j=1}^N a_{ij} I_j
\eq
More generally, let us denote by $\vec{I} = \left(I_1,...,I_N\right)$ the vector of the master integrals,
and by $\vec{x} = \left(x_1,...,x_n\right)$ the vector of kinematic variables the master integrals depend on.
Repeating the above procedure for every master integral and every kinematic variable we 
obtain a system of differential equations of Fuchsian type
\bq
\label{diff_eq}
 d\vec{I} & = & A \vec{I},
\eq
where $A$ is a matrix-valued one-form
\bq
 A & = & 
 \sum\limits_{i=1}^n A_i dx_i.
\eq
The matrix-valued one-form $A$ satisfies the integrability condition $dA - A \wedge A = 0$.

There is a class of Feynman integrals, which may be expressed in terms of multiple polylogarithms.
Multiple polylogarithms are defined by the nested sum \cite{Goncharov_no_note,Goncharov:2001,Borwein,Moch:2001zr}
\bq
 \mbox{Li}_{m_1,m_2,...,m_k}(x_1,x_2,...,x_k) & = & 
 \sum\limits_{n_1 > n_2 > ... > n_k > 0}^\infty 
 \;\;\;
 \frac{x_1^{n_1}}{n_1^{m_1}} \cdot \frac{x_2^{n_2}}{n_2^{m_2}} \cdot ... \cdot \frac{x_k^{n_k}}{n_k^{m_k}}.
\eq
There is an alternative definition based on iterated integrals
\bq
G(z_1,...,z_k;y) & = & \int\limits_0^y \frac{dt_1}{t_1-z_1}
 \int\limits_0^{t_1} \frac{dt_2}{t_2-z_2} ...
 \int\limits_0^{t_{k-1}} \frac{dt_k}{t_k-z_k}.
\eq
The two notations are related by
\bq
\mbox{Li}_{m_1,...,m_k}(x_1,...,x_k) 
 & = &
 (-1)^k 
 G_{m_1,...,m_k}\left( \frac{1}{x_1}, \frac{1}{x_1 x_2}, ..., \frac{1}{x_1...x_k};1 \right),
\eq
where
\bq
G_{m_1,...,m_k}(z_1,...,z_k;y) & = &
 G(\underbrace{0,...,0}_{m_1-1},z_1,...,z_{k-1},\underbrace{0...,0}_{m_k-1},z_k;y).
\eq
Let us return to the differential equation~(\ref{diff_eq}).
If we change the basis of the master integrals $\vec{J} = U \vec{I}$, the differential equation becomes
\bq
 d \vec{J} = A' \vec{J},
 & & A' = U A U^{-1} - U d U^{-1}.
\eq
Suppose further one finds a transformation matrix $U$, such that
\bq
 A' & = & \eps \sum\limits_j \; C_j \; d\ln p_j(\vec{x}),
\eq
where the dimensional regularisation parameter $\eps$ appears only as prefactor,
the $C_j$ are matrices with constant entries, and where the 
$p_j(\vec{x})$ are polynomials in the external variables,
then the system of differential equations is easily solved in terms of multiple polylogarithms \cite{Henn:2013pwa,Henn:2014qga}.
In order to obtain the $\eps$-form we may
perform a rational or algebraic transformation of the kinematic variables
\bq
 (x_1,...,x_n) & \rightarrow & (x_1',...,x_n').
\eq
This corresponds to a change of variables in the base manifold.
A change of kinematic variables is often done to absorb square roots for massive integrals.
In addition, we may change the basis of the master integrals
\bq
 \vec{I} & \rightarrow & U \vec{I},
\eq
where $U$ is rational in the kinematic variables.
This corresponds to a change of basis in the fibre.
Methods to find the right transformation have been discussed in \cite{Gehrmann:2014bfa,Argeri:2014qva,Lee:2014ioa,Prausa:2017ltv,Gituliar:2017vzm,Meyer:2016slj,Adams:2017tga,Lee:2017oca,Meyer:2017joq,Becchetti:2017abb}.

At the end of the day we would like to evaluate the multiple polylogarithms numerically, taking into account that 
the multiple polylogarithms $\mbox{Li}_{m_1,m_2,...,,m_k}(x_1,x_2,...,x_k)$ have branch cuts
as a function of the $k$ complex variables $x_1$, $x_2$, ..., $x_k$.
The numerical evaluation can be done as follows:
One uses a truncation of the sum representation within the region of convergence.
The integral representation is used to map the arguments into the region of convergence.
On top of that, acceleration techniques are used to speed up the computation \cite{Vollinga:2004sn}.

% -----------------------------------------------------------------------------

\section{Beyond multiple polylogarithms: Single scale integrals}

Starting from two-loops, there are integrals which cannot be expressed in terms of multiple polylogarithms.
The simplest example is given by the 
two-loop sunrise integral \cite{Broadhurst:1993mw,Berends:1993ee,Bauberger:1994nk,Bauberger:1994by,Bauberger:1994hx,Caffo:1998du,Laporta:2004rb,Kniehl:2005bc,Groote:2005ay,Groote:2012pa,Bailey:2008ib,MullerStach:2011ru,Adams:2013nia,Bloch:2013tra,Adams:2014vja,Adams:2015gva,Adams:2015ydq,Remiddi:2013joa,Bloch:2016izu}
with equal masses.
A slightly more complicated integral is the two-loop kite integral \cite{Sabry:1962,Remiddi:2016gno,Adams:2016xah,Adams:2017ejb,Bogner:2017vim}, which contains the sunrise integral as a sub-topology.
Both integrals depend on a single dimensionless variable $t/m^2$.
In the following we will change the variable from $t/m^2$ to the nome $q$ of an elliptic curve or the parameter $\tau$, related
to the nome by $q=e^{i \pi \tau}$.
Before giving a definition of these new variables, let us first see how an elliptic curve emerges.
For the sunrise integral there are two possibilities.
The first option reads off an elliptic curve from 
the Feynman graph polynomial
\bq
\label{elliptic_curve}
 E_{\mathrm{graph}}
 & : &
 - x_1 x_2 x_3 t + m^2 \left( x_1 + x_2 + x_3 \right) \left( x_1 x_2 + x_2 x_3 + x_3 x_1 \right) 
 \; = \; 0,
\eq
the second option obtains an elliptic curve from the maximal cut \cite{Baikov:1996iu,Lee:2009dh,Kosower:2011ty,CaronHuot:2012ab,Frellesvig:2017aai,Bosma:2017ens,Harley:2017qut}
of the sunrise integral
\bq
 E_{\mathrm{cut}}
 & : &
 y^2 
 -
 \left(x - \frac{t}{m^2} \right) 
 \left(x + 4 - \frac{t}{m^2} \right) 
 \left(x^2 + 2 x + 1 - 4 \frac{t}{m^2} \right)
 \; = \; 0.
\eq
In the following we will consider the elliptic curve of eq.~(\ref{elliptic_curve}).
The periods $\psi_1$, $\psi_2$ of the elliptic curve are solutions of the homogeneous differential equation \cite{Adams:2013nia}.
In general, the maximal cut of a Feynman integral is a solution of the homogeneous differential equation
for this Feynman integral \cite{Primo:2016ebd}.
We define the new variables $\tau$ and $q$ by
\bq
 \tau \; = \; \frac{\psi_2}{\psi_1},
 & &
 q \; = \; e^{i \pi \tau}.
\eq
Let us now turn to the transcendental functions, in which we may express the sunrise and the kite integral.
We remind the reader of the definition of the classical polylogarithms
\bq
 \mathrm{Li}_n\left(x\right) & = & \sum\limits_{j=1}^\infty \; \frac{x^j}{j^n}.
\eq
Starting from this expression, we consider a generalisation with two sums, which
are coupled through the variable $q$:
\bq
 \mathrm{ELi}_{n;m}\left(x;y;q\right) & = & 
 \sum\limits_{j=1}^\infty \sum\limits_{k=1}^\infty \; \frac{x^j}{j^n} \frac{y^k}{k^m} q^{j k}.
\eq
The elliptic dilogarithm is a linear combination of these functions and the classical dilogarithm:
\bq
 \mathrm{E}_{2;0}\left(x;y;q\right)
 & = &
 \frac{1}{i}
 \left[
 \frac{1}{2} \mathrm{Li}_2\left( x \right) 
 - \frac{1}{2} \mathrm{Li}_2\left( x^{-1} \right)
 + \mathrm{ELi}_{2;0}\left(x;y;q\right)
 - \mathrm{ELi}_{2;0}\left(x^{-1};y^{-1};q\right)
 \right].
\eq
In the mathematical literature there exist various slightly different definitions 
of elliptic polylogarithms \cite{Beilinson:1994,Levin:1997,Levin:2007,Enriquez:2010,Brown:2011,Wildeshaus,Bloch:2013tra,Bloch:2014qca,Remiddi:2017har}.
In order to express the sunrise and the kite integral to all orders in $\eps$ we introduce the functions
\bq
\label{def_ELi}
\lefteqn{
 \mathrm{ELi}_{n_1,...,n_l;m_1,...,m_l;2o_1,...,2o_{l-1}}\left(x_1,...,x_l;y_1,...,y_l;q\right)
 = }
 & & \nonumber \\
 & = &
 \sum\limits_{j_1=1}^\infty ... \sum\limits_{j_l=1}^\infty
 \sum\limits_{k_1=1}^\infty ... \sum\limits_{k_l=1}^\infty
 \;\;
 \frac{x_1^{j_1}}{j_1^{n_1}} ... \frac{x_l^{j_l}}{j_l^{n_l}}
 \;\;
 \frac{y_1^{k_1}}{k_1^{m_1}} ... \frac{y_l^{k_l}}{k_l^{m_l}}
 \;\;
 \frac{q^{j_1 k_1 + ... + j_l k_l}}{\prod\limits_{i=1}^{l-1} \left(j_i k_i + ... + j_l k_l \right)^{o_i}}.
\eq
Let us write the Taylor expansion of the sunrise integral around $D=2-2\eps$ as
\bq
 S & = &
 \frac{\psi_1}{\pi}
 \sum\limits_{j=0}^\infty \eps^j E^{(j)}.
\eq
Each term in this $\eps$-series is 
of the form
\bq
 E^{(j)} & \sim &
 \mbox{linear combination of} \;
 \mathrm{ELi}_{n_1,...,n_l;m_1,...,m_l;2o_1,...,2o_{l-1}}
 \;\; \mbox{and} \;\; \mathrm{Li}_{n_1,...,n_l}.
\eq
Using dimensional-shift relations this translates to the expansion around $D=4-2\eps$.
Thus we find that the functions of eq.~(\ref{def_ELi}) together with the multiple polylogarithms 
are the class of functions to express the equal mass sunrise graph and the kite integral to all orders in $\eps$ \cite{Adams:2015ydq,Adams:2016xah}.

The functions in eq.~(\ref{def_ELi}) are defined as multiple sums. We may ask if every term 
in the $\eps$-expansion can be expressed in terms of iterated integrals.
For the equal-mass sunrise integral and the kite integral this is indeed the case and relates these Feynman integrals to modular forms \cite{Adams:2017ejb}.
A function $f(\tau)$ on the complex upper half plane is a modular form of weight $k$ for $\mathrm{SL}_2(\mathbb{Z}$) if
$f$ transforms under M\"obius transformations as
\bq
 f\left( \frac{a\tau+b}{c\tau+d} \right) = (c\tau+d)^k \cdot f(\tau) 
 \qquad \mbox{for} \;\; \left( \begin{array}{cc}
a & b \\ 
c & d
\end{array} \right) \in \mathrm{SL}_2(\mathbb{Z}).
\eq
In addition, $f$ is required to be holomorphic on the complex upper half plane and at $\tau=i \infty$.
Furthermore, there are modular forms for congruence subgroups of $\mathrm{SL}_2(\mathbb{Z})$.
We introduce iterated integrals of modular forms
\bq
 I\left(f_1,f_2,...,f_n;\bar{q}\right)
 & = &
 \left(2 \pi i \right)^n
 \int\limits_{\tau_0}^{\tau} d\tau_1
 f_1\left(\tau_1\right)
 \int\limits_{\tau_0}^{\tau_1} d\tau_2
 f_2\left(\tau_2\right)
 ...
 \int\limits_{\tau_0}^{\tau_{n-1}} d\tau_n
 f_n\left(\tau_n\right),
 \;\;\;\;\;\;
 \bar{q} = e^{2\pi i \tau}.
\eq
As base point it is convenient to take $\tau_0=i \infty$.
Repeated sequences of letters are abbreviated as in $\{f_{1},f_{2}\}^3 = f_{1},f_{2},f_{1},f_{2},f_{1},f_{2}$.
With the help of the iterated integrals of modular forms one finds a compact all-order expression for the 
equal-mass sunrise integral around $D=2-2\eps$ dimensions:
\bq
 S 
 & = & 
 \frac{\psi_1}{\pi}
 e^{-\eps I(f_2;q) + 2\sum\limits_{n=2}^\infty \frac{\left(-1\right)^n}{n} \zeta_n \eps^n} 
 \left\{
  \sum\limits_{j=0}^\infty 
   \eps^j 
   \sum\limits_{k=0}^{\lfloor \frac{j}{2} \rfloor} I\left( \left\{1,f_4\right\}^k, 1, f_3, \left\{f_2\right\}^{j-2k}; q\right)
 \right.
 \nonumber \\
 & &
 \left.
 +
 \left[
 \sum\limits_{j=0}^\infty 
 \left(
 \eps^{2j} I\left(\left\{1,f_4\right\}^j;q \right)
 -
 \frac{1}{2} \eps^{2j+1} I\left(\left\{1,f_4\right\}^j,1;q \right)
 \right)
 \right]
 \sum\limits_{k=0}^\infty \eps^k B^{(k)}
 \right\},
\eq
where the $B^{(k)}$'s are boundary constants.
This expression has uniform depth, i.e. at order $\eps^j$ one has exactly $(j+2)$ iterated integrations.
The alphabet is given by four modular forms $1$, $f_2$, $f_3$, $f_4$.
To give an example, 
the modular form $f_3$ is given by
\bq
 f_3 
 & = &
 -
 \frac{1}{24}
 \left( \frac{\psi_1}{\pi} \right)^3
 \;
 \frac{t\left( t - m^2 \right)\left( t - 9 m^2 \right)}{m^6}.
\eq
$f_3$ may be expressed as a linear combination of generalised Eisenstein series, which makes the property of being a modular form manifest.

Let us now return to question if there is an $\eps$-form for the differential equations for the sunrise and kite integrals.
It is not possible to obtain an $\eps$-form by an algebraic change of variables
and/or an algebraic transformation of the basis of master integrals.
However by the (non-algebraic) change of variables from $t$ to $\tau$ and by factoring off
the (non-algebraic) expression $\psi_1/\pi$ from the master integrals in the sunrise sector
one obtains an $\eps$-form for the kite/sunrise family:
\bq
 \frac{d}{d\tau} \vec{I}
 & = &
 \eps \; A(\tau) \; \vec{I},
\eq
where $A(\tau)$ is an $\eps$-independent $8 \times 8$-matrix whose entries are modular forms.

Let us turn to the numerical evaluation:
The complete elliptic integrals entering $\psi_1$ can be computed efficiently from the arithmetic-geometric mean.
The numerical evaluation of the ${\mathrm{ELi}}$-functions is straightforward in the region where the sum converges:
One simply truncates the $q$-series at a certain order such that the desired numerical precision is reached.
Methods to map the arguments outside the region of convergence into this region have been discussed in
\cite{Passarino:2017EPJC}.
It turns out that for the sunrise integral and the kite integral the
$q$-series converges for all $t \in {\mathbb R}\backslash \{m^2, 9m^2, \infty\}$,
in particular there is no need to distinguish the cases $t<0$, $0<t<m^2$, $m^2<t<9m^2$ or $9m^2<t$ \cite{Bogner:2017vim}.

% -----------------------------------------------------------------------------

\section{Towards multi-scale integrals beyond multiple polylogarithms}

Let us now turn from single-scale integrals to multi-scale integrals.
We are interested in the ones, which are not expressible in 
in terms of multiple polylogarithms \cite{Adams:2014vja,Adams:2015gva,Bonciani:2016qxi,vonManteuffel:2017hms,Ablinger:2017bjx,Bourjaily:2017bsb,Hidding:2017jkk},
but are expressible in terms of elliptic generalisations of these functions.
Therefore we expect in the differential equation for a given master integral irreducible second-order factors.
A system of first-oder differential equations is easily converted to a higher-order differential equation
for a single master integral.
We may work modulo sub-topologies, therefore the order of the differential equation is given by the number
of master integrals in this sector.
The number of master integrals in a given sector may be larger than $2$ and we face the question on how to
transform to a suitable basis of master integrals, which decouples the original system of differential equations
at order $\eps^0$ to a system of maximal block size of $2$.
This can be done by exploiting the factorisation properties of the Picard-Fuchs operator\cite{Adams:2017tga}.
To this aim one first projects the problem to a single-scale problem by setting
$x_i\left(\lambda\right) = \alpha_i \lambda$ with 
$\alpha=[\alpha_1:...:\alpha_n] \in {\mathbb C} {\mathbb P}^{n-1}$
and by viewing the master integrals as functions of $\lambda$.
For the derivative with respect to $\lambda$ we have
\bq
 \frac{d}{d\lambda} \vec{I}
 & = &
 B \vec{I},
 \;\;\;\;\;\;
 B \; = \;
 \sum\limits_{i=1}^n \alpha_i A_i,
 \;\;\;\;\;\;
 B \; = \; B^{(0)} + \sum\limits_{j>0} \eps^j B^{(j)}.
\eq
In order to find the required transformation we may work modulo $\eps$-corrections, i.e. 
we focus on $B^{(0)}$.
Let $I$ be one of the master integrals $\{I_1,...,I_N\}$.
We determine the largest number $r$, such that the matrix which expresses 
$I$, $(d/d\lambda)I$, ..., $(d/d\lambda)^{r-1}I$ in terms of the original set $\{I_1,...,I_N\}$ has full rank.
It follows that $(d/d\lambda)^rI$ can be written as a linear combination of $I, ..., (d/d\lambda)^{r-1}I$.
This defines the Picard-Fuchs operator $L_r$ for the master integral $I$ with respect to $\lambda$:
\bq
 L_{r} I & = & 0,
 \;\;\;\;\;\;
 L_r \; = \; \sum\limits_{k=0}^r R_k \frac{d^k}{d\lambda^k}.
\eq
$L_r$ is easily found by transforming to a basis which contains $I, ..., (d/d\lambda)^{r-1}I$.
We may factor the differential operator into irreducible factors \cite{vanHoeij:1997}.
\bq
 L_r
 & = &
 L_{1,r_1} L_{2,r_2} ... L_{s,r_s}, 
\eq
where $L_{i,r_i}$ denotes a differential operator of order $r_i$.
We may then convert the system of differential equations
at order $\eps^0$ into a block triangular form
with blocks of size $r_1$, $r_2$, ..., $r_s$.
A basis for block $i$ is given by
\bq
 J_{i,j} & = &
 \frac{d^{j-1}}{d\lambda^{j-1}} L_{i+1,r_{i+1}} ... L_{s,r_s} I,
 \;\;\;\; 1 \le j \le r_i.
 \;\;\;
\eq
This decouples the original system into sub-systems of size $r_1$, $r_2$, ..., $r_s$.
Let us write the transformation to the new basis as
$\vec{J} = V\left(\alpha_1,...,\alpha_{n-1},\lambda\right) \vec{I}$.
Setting 
\bq
 U & = & V\left(\frac{x_1}{x_n},...,\frac{x_{n-1}}{x_n},x_n\right)
\eq
gives a transformation in terms of the original variables $x_1$, ..., $x_n$.
Terms in the original matrix $A$ of the form $d \ln Z(x_1,...,x_n)$,
where $Z(x_1,...,x_n)$ is a rational function in $(x_1,...,x_n)$ and homogeneous of degree zero in $(x_1,...,x_n)$,
map to zero in the matrix $B$.
These terms are in many cases easily removed by a subsequent transformation.
Let us look at an example. For the planar double-box integral for $t\bar{t}$-production with a closed top loop
one finds in the top sector five master integrals.
These may be decoupled as
\bq
 5 & = & 1 + 2 + 1 + 1.
\eq
Thus we need to solve only two coupled equations, not five.

% -----------------------------------------------------------------------------

\section{Conclusions}

Differential equations are a powerful tool to compute Feynman integrals.
If a system can be transformed to an $\eps$-form with rational or algebraic transformations, 
a solution in terms of multiple polylogarithms is easily obtained.
There are however systems, where within rational transformations at order $\eps^0$ two coupled equations remain.
The simplest examples of these are the Feynman integrals belonging to the families of the
equal-mass sunrise integral and the kite integral.
They evaluate to elliptic generalisations of multiple polylogarithms. 
The iterated integral representation is given in terms of iterated integrals of modular forms.
With a non-algebraic change of variables and a non-algebraic basis transformation it is possible to obtain
an $\eps$-form.
For Feynman integrals depending on several variables 
the factorisation properties of the Picard-Fuchs operator allows us to find the irreducible blocks.

% -----------------------------------------------------------------------------
% references
\bibliography{/home/stefanw/notes/biblio}
\bibliographystyle{/home/stefanw/latex-style/h-physrev5}

\end{document}